\documentclass[alp,jap,twocolumn,showpacs,superscriptaddress,groupedaddress]{revtex4}  
\usepackage{graphicx}  
\usepackage{dcolumn}   
\usepackage{bm}        
\usepackage{amssymb}   
\usepackage{amsmath}

\begin{document}


\title{Coherent Manipulation of Thermal Transport by Tunable Electron-Photon and Electron-Phonon Interaction}
\author {Federico Paolucci}
\email{federico.paolucci@nano.cnr.it}
\affiliation{ NEST, Instituto Nanoscienze-CNR and Scuola Normale Superiore, I-56127 Pisa, Italy}
\author {Giuliano Timossi}
\affiliation{ NEST, Instituto Nanoscienze-CNR and Scuola Normale Superiore, I-56127 Pisa, Italy}
\author {Paolo Solinas}
\affiliation{ SPIN-CNR, Via Dodecaneso 33, I-16146 Genova, Italy}
\author {Francesco Giazotto}
\email{francesco.giazotto@sns.it}
\affiliation{ NEST, Instituto Nanoscienze-CNR and Scuola Normale Superiore, I-56127 Pisa, Italy}

\begin{abstract}
We propose a system where coherent thermal transport between two reservoirs in non-galvanic contact is modulated by independently tuning the electron-photon and the electron-phonon coupling. The scheme is based on two gate-controlled electrodes capacitively coupled through a dc-SQUID as intermediate phase-tunable resonator. Thereby the electron-photon interaction is modulated by controlling the flux threading the dc-SQUID and the impedance of the two reservoirs, while the electron-phonon coupling is tuned by controlling the charge carrier concentration in the electrodes. To quantitatively evaluate the behavior of the system we propose to exploit graphene reservoirs. In this case, the scheme can work at temperatures reaching $1~$K, with unprecedented temperature modulations as large as $245~$mK, transmittance up to $99\%$ and energy conversion efficiency up to $50\%$. Finally, the accuracy of heat transport control allows us to use this system as an experimental tool to determine the electron-phonon coupling in two dimensional electronic systems (2DES).
\end{abstract}

\maketitle

\section{Introduction}

\begin{figure}
\includegraphics {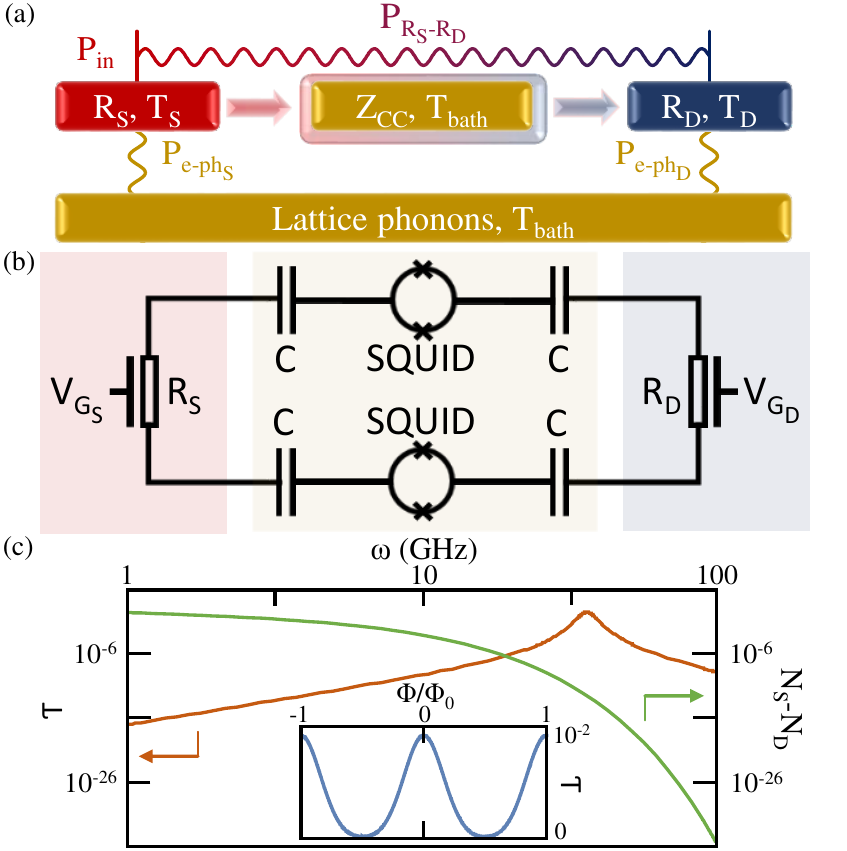}
\caption{\label{fig:Fig1} (a) Thermal model of the system: the red and blue rectangles represent the source ($S$) and drain ($D$) electrode characterized by different impedances ($R_S$ and $R_D$) and temperatures ($T_S$ and $T_D$). The gold rectangle depicts the coupling circuit $CC$ of impedance $Z_{CC}$ at temperature $T_{bath}$. The wavy lines portray the heat exchange involved: $P_{e-ph_{S}}$  and $P_{e-ph_{D}}$ represent the power losses due to electron-phonon coupling in the reservoirs, while  $P_{R_{S}-R_{D}}$ is power transmitted from $S$ to $D$ due to the electron-photon interaction mediated by $CC$. (b) Equivalent electric circuit of the system: $S$ and $D$ electrodes are represented as resistors $R_S$ and $R_D$ controlled by the gate voltages $V_{G_S}$ and $V_{G_D}$, respectively. The four capacitors $C$ and the two SQUIDs composing the $CC$ are shown. (c) Typical transfer function $\mathcal{T}(\omega)$ (orange line) and difference between bosonic functions for the photons in $S$ and $D$ (green line) as a function of frequency for a flux $\Phi=0$. Inset: $\mathcal{T}(\omega=50~$GHz$)$ as a function of $\Phi$. The used parameters are: $C_{SQUID}=29~$fF, $L_0=1~$nH, $C=750~$fF and $R_S=R_D=50~\Omega$.}
\end{figure}

Control and manipulation of thermal currents in solid-state structures is of particular interest especially at the nanoscale where heat strongly affects the physical properties of the systems. In this direction, coherent caloritronics \cite{Meschke2006, Giazotto2012, Martinez2014b}, which takes advantage of phase-coherent mastering the heat current in solid-state nanostructures, represents a crucial breakthrough in several fields of science at cryogenic temperatures such as quantum computing \cite{Spilla2014}, ultrasensitive radiation detectors \cite{Giazotto2008} and electron cooling \cite{Quaranta2011, Solinas2016}. Although phase coherence plays a fundamental role in the functionalities of several nano-electronic devices, the impact of coherence in caloritronics is far to be completely understood. Despite it is smaller than galvanic thermal transport \cite{Bosisio2016}, electron-photon mediated heat transfer \cite{Ojanen2008, Pascal2011, Meschke2006, Partanen2016} provides the possibility of contactless heating or cooling bodies in non-galvanic contact allowing to investigate low energy physics in small quantum devices, and put the basis of novel-concept logic elements. These approaches assume that at low temperatures the phonon modes become effectively frozen \cite{Meschke2006, Schmidt2004} and the energy losses through electron-phonon coupling are minimized. 

Yet, the possibility of in-situ tuning the impedance matching and the electron-phonon coupling in such systems can pave the way to new device concepts, unexplored physical effects and novel quantum-state engineering. Here, we propose a device able to fully control the coherent heat transport between two bodies in non-galvanic contact by modulating their electron-photon interaction and their electron-phonon coupling. 

\section{Thermal Model}
The system we study consists of a source $S$ and a drain $D$ electron reservoirs (see  Fig.~\ref{fig:Fig1}-a) interacting through an intermediate coupling circuit ($CC$). The lattice phonons residing in every element of the structure are assumed to be thermalized with the substrate phonons at bath temperature $T_{bath}$, because of the vanishing Kapitza resistance between thin conductors and the substrate at low temperatures \cite{Giazotto2012, Wellstood1994}. The heat losses are exclusively due to the electron-phonon interaction. In the steady state, a power $P_{in}$ injected into the source electrode originates a heat flow described by the following system of energy balance equations:
\begin{equation}
\begin{split}
\begin{cases}
P_{in}=P_{e-ph_{S}}+P_{R_S-R_D}\\
P_{R_S-R_D}=P_{e-ph_{D}}.\\
\end{cases}
\end{split}
\label{eq:Pbalance}
\end{equation}
Above $P_{e-ph_{S}}$ and $P_{e-ph_{D}}$ are the power losses due to electron-phonon coupling in $S$ and $D$, respectively, and $P_{R_S-R_D}$ is the electron-photon mediated power transfer between the two reservoirs in non-galvanic contact. As a consequence, for a non-zero $P_{R_S-R_D}$ the temperatures of the source $T_S$ and drain $T_D$ electrodes differ from the phonon temperature $T_{bath}$ and follow $T_S\geq T_D\geq T_{bath}$ (in the case of positive $P_{in}$).

Here, we focus initially our attention on $P_{R_{S}-R_{D}}$ . The heat transport between two remote bodies mediated by the electron-photon interaction has been studied within a nonequilibrium Green's function formalism \cite{Ojanen2008} or a circuital approach \cite{Pascal2011}. For simplicity, our analysis is based on the circuital approach, since the two methods give equivalent results \cite{Pascal2011}. A thermal excitation generates a fluctuating noise current in each element of the system. The thermal current between $S$ and $D$ can be calculated as the difference between the power emitted from the drain and the source reservoirs due to the electromagnetic noise, and it can be expressed as:
\begin{equation}
\begin{split}
P_{R_S-R_D}=\int_0^\infty\frac{d\omega}{2\pi}\hbar\omega\mathcal{T}(\omega)\left[N_D(\omega)-N_S(\omega)\right],
\end{split}
\label{eq:PSD}
\end{equation}
where  $\mathcal{T}(\omega)$ is the photonic transfer function of the system, while $N_S$ and $N_D$ are the Bose-Einstein distributions of the source and drain photon reservoirs, respectively. The frequency-dependent transfer function is defined as \cite{Pascal2011}:

\begin{equation}
\begin{split}
\mathcal{T}(\omega)=\frac{4\Re[Z_S(\omega)]\Re[Z_D(\omega)]}{|Z_{TOT}(\omega)|^2}=\\
=\frac{4R_SR_D}{|R_S+Z_{CC}(\omega)+R_D|^2},
\end{split}
\label{eq:Transf}
\end{equation}
where $Z_S$, $Z_D$ and $Z_{CC}(\omega)$ are the frequency-dependent impedances of $S$, $D$ and coupling circuit, respectively. The impedance of source and drain are assumed to be completely resistive ($Z_i=R_i$ with $i=S,D$). In order to phase coherently modulate the electron-photon interaction, it is necessary to design an appropriate electronic circuit with a tunable $Z_{CC}$. A typical intermediate quantum circuit is represented by a flux-controlled dc-SQUID (superconducting quantum interference device)\cite{Meschke2006, Ojanen2008, Pascal2011, Giazotto2006},  which is modeled as a $LC$ resonator with a magnetic flux-dependent inductance $L_{SQUID}(\Phi)=L_0/|\cos(\pi\Phi/\Phi_0)|$\cite{Meschke2006, Bosisio2016}, where $L_0 \propto 1/I_C$ is the Josephson inductance arising from the device critical current $I_C$, $\Phi$ is the flux threading the loop and $\Phi_0=2.067\times10^{-15}~$Wb is the flux quantum. Therefore, the SQUID acts as a phase-dependent  thermal modulator. 

The implementation that we have chosen in order to ensure phase-coherent heat modulation in non-galvanic contacts is depicted in Fig.~\ref{fig:Fig1}-b, where a dc-SQUID is capacitively-coupled with the electron reservoirs. Then the total series impedance of the coupling circuit is $Z_{CC}=2Z_C+Z_{SQUID}$, where $Z_C=1/i\omega C$ and $Z_{SQUID}=i 2 \omega L_{SQUID}/\left[1-\left(\omega\sqrt{L_{SQUID}C_{SQUID}}\right)^{2}\right]$ \cite{Meschke2006} are the impedances of the capacitor and the dc-SQUID, respectively. Contrary to the narrow-band pass filter characteristic of an inductive coupling \cite{Ojanen2008, Pascal2011}, the broad-band pass filter behavior of capacitive coupling ensures the maximum power transfer across the device, because the term $[N_D(\omega)-N_S(\omega)]$ of Eq.~(\ref{eq:PSD}) shows the maximum value at low photon frequency (see Fig.~\ref{fig:Fig1}-c). Furthermore, thermal transport across the system $P_{R_{S}-R_{D}}$ is maximized by increasing the mutual inductance $M$ between the reservoirs and the coupling circuit \cite{Ojanen2008, Pascal2011}, but the modulation of $P_{R_{S}-R_{D}}$ is maximum when the geometrical inductance of $CC$ is low and the Josephson inductance of the SQUID dominates. The impossibility to simultaneously fulfill these two conditions makes inductive coupling unusable in real caloritronic devices. Oppositely, large capacitances can be easily realized in experiments and do not require any constrain in the SQUID geometry. The reactive impedances due to the presence of the gates coupled to the source and drain electrodes are much smaller than the coupling capacitances and, therefore, they can be neglected in the circuit model of the device. 

Equation~(\ref{eq:Transf}) clearly shows that the transfer function $\mathcal{T}(\omega)$ also depends on the values of the source and drain impedances, and it is maximized in the case of a matched circuit ($Z_S=Z_D$). Therefore, the electron-photon interaction can also be modulated by controlling the value of $R_S$ and $R_D$. We propose the use of a two dimensional electron system (2DES) as material implementing the $S$ and $D$ electrodes. By employing two gates ($V_{G_S}$ and $V_{G_D}$ in Fig.~\ref{fig:Fig1}-b) it is possible to independently control the charge carrier concentration $n$ of the two reservoirs. As a consequence, the modulation of their resistance ($R_S,~R_D\propto1/ n$) is reflected in a change of $P_{R_S-R_D}$ and thereby in thermal transport across the system. The parasitic capacitance associated to the presence of the gate electrodes is negligible small compared to the coupling capacitance $C$, therefore the impedance of $S$ and $D$ can be still considered fully resistive ($Z_i=R_i$ with $i=S,D$) and Eq.~(\ref{eq:Transf}) holds. This architecture represents a perfect platform to study the impact of impedance matching on the thermal propoerties of solid-state nano-structures. For instance, a variable impedance source electrode can be beneficial to control the temperature of a quantum device (as an example a quantum dot) without interfering with the behavior of the latter through contactless heat transport. 

We now turn to the analysis of $P_{e-ph}$. In the following we impose a constant $T_S$, thereby we will consider only the impact of $P_{e-ph_{D}}$ on thermal transport. The changes in the electron-phonon coupling in the source electrodes $P_{e-ph_{D}}$ has impact only on the power that is necessary to provide to the system in order to reach the desired value of $T_S$. Equation~(\ref{eq:Pbalance}) shows that the thermal efficiency of the system strongly depends on the electron-phonon coupling of the source and drain electrodes. At temperatures lower than the Debye temperature, the power is dissipated to the lattice only thorough the acoustic phonons. In a clean metal the electron-phonon dissipation takes the form $P_{e-ph}=\Sigma V(T_e^5-T_{bath}^5)$, where $\Sigma$ is the electron-phonon coupling constant and $V$ is the volume \cite{Giazotto2006}. The coupling constant is peculiar for every material and it is defined as $\Sigma = 12\zeta(5)k_B^5|D_P|^2/\pi\hbar^5v_S^3$ where $\zeta(5)\approx 1.0369$, $D_P$ is the deformation potential and $v_S$ is the speed of sound in the material. In the dirty limit (diffusive regime) $P_{e-ph}$ scales with $T^4$ or $T^6$ depending on the nature of disorder \cite{Sergeev2000, Karvonen2004}. In conventional caloritronic devices the reservoirs are usually made of metals, and $\Sigma$ is fixed by the choice of material.  As mentioned above, we propose to use gated 2DESs as source and drain electrodes. In two dimensional electronic systems the electron phonon coupling constant depends on the charge carrier concentration \cite{Gasparinetti2011, Giazotto2006b}. Therefore, our system allows to in-situ modulate $\Sigma$  and, therefore, the temperature of the drain electrode $T_D$ by tuning the gate voltage applied to the reservoirs. This new knob increases the versatility of caloritronic devices paving the way to novel applications.

\begin{figure}
\includegraphics {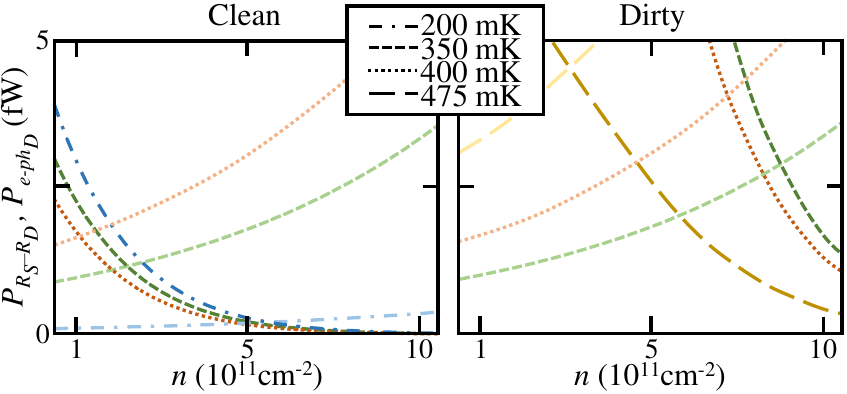}
\caption{\label{fig:Fig2mod1} $P_{R_S-R_D}$ at $\Phi=0$ (dark) and $P_{e-ph_{D}}$ (light) as a function of $n=n_S=n_D$ for different values of $T_S$ in the limit of clean (left) and dirty graphene (right). }
\end{figure}

In order to quantify the impact of $P_{R_S-R_D}$ and  $P_{e-ph_{D}}$ on the modulation of the heat transfer we suppose to exploit $S$ and $D$ electrodes made of graphene. The thermal properties of graphene have been extensively studied \cite{Chae2010, Betz2012, Johannsen2013, Laitinen2013}, but it has never been used in coherent caloritronic systems so far. Its small volume, versatility and the possibility of producing samples of different quality make graphene the prototype candidate of tunable material for caloritronic applications. The power exchanged between electrons and phonons in graphene depends on its charge carrier concentration, which can be controlled through gate electrodes, and mobility \cite{CastroNeto2009}. 

In the clean limit (i.e., implying the mean free path $\ell \geq 1\mu m$) it takes the form \cite{McKitterick2016}:

\begin{equation}
\begin{split}
P_{e-ph_{clean}}=A\Sigma_{clean}\left( T_{e}^4-T_{bath}^4\right),
\end{split}
\label{eq:Pclean}
\end{equation}
where $A$ is the area of the graphene sheet, $T_{e}$ is the electronic temperature and $\Sigma_{clean}=(\pi^2D_P^2|E_F|k_B^4)/(15\rho_M\hbar^5v_F^3v_S^3)$ is the electron-phonon coupling constant. In the latter, $v_F=10^6$m$/$s is the Fermi velocity, $D_P$ is the deformation potential, $E_F=\hbar v_F \sqrt{\pi n}$ is the Fermi energy, $\rho_M$ is the mass density and $v_S=2\times10^{14}$m$/$s is the speed of sound in graphene. The temperature of lattice phonons of graphene is assumed to be the same of the substrate $T_{bath}$, because the vanishing Kapitza resistance ensures full thermalization at low temperatures \cite{Balandin2011}. In the dirty limit ($\ell \leq 100~$nm), the power losses due to electron-phonon coupling have a $T^3$ dependence and are expressed \cite{Chen2012} as follows:
\begin{equation}
\begin{split}
P_{e-ph_{dirty}}=A\Sigma_{dirty}\left( T_{e}^3-T_{bath}^3\right),
\end{split}
\label{eq:Pdirty}
\end{equation}
where the electron-phonon coupling constant takes the form $\Sigma_{dirty}=(1.2D_P^2|E_F|k_B^3)/(\pi^2\rho_M\hbar^4v_F^3v_S^2\ell)$. 

\section{Results and discussion}

In order to solve the energy balance system [Eq.~(\ref{eq:Pbalance})] and quantitatively analyze the behavior of the efficiency of the system, we use the following values for the physical quantities: $T_{bath}=10~$mK, $C_{SQUID}=29~$fF, $L_0=1~$nH, $C=750~$fF and $A=12.5~\mu$m$^2$. Since the transfer function is maximized for a matched circuit, i.e., $R_S=R_D$, in the following we assume $n=n_S=n_D$ where $n_S$ and $n_D$ are the charge concentrations in $S$ and $D$, respectively. Figure~\ref{fig:Fig2mod1} shows $P_{R_S-R_D}$ (dark lines) and $P_{e-ph_{D}}$ (light lines) as a function of the carrier concentration $n$ for different values of $T_S$. Differently from other 2DES  \cite{Gasparinetti2011, Giazotto2006b}, in graphene the electron-phonon coupling constant $\Sigma \propto \sqrt{n}$. As a consequence, the power adsorbed by phonons increases monotonically with source temperature and carrier concentration both in the clean and dirty limit (see Fig. \ref{fig:Fig2mod1}).  Notably, $P_{e-ph_{D}}$ is always of the order of fW for our system. In metals, the electron phonon coupling is typically stronger and thermal losses are larger. For instance, at $T=200~$mK in our system  $P_{e-ph_{D}}\leq 1fW$, while AlMn or Cu thin films of comparable dimensions show a dissipated power of $\sim30~$fW \cite{Martinez2014} or $\sim100~$fW \cite{Giazotto2012}, respectively. On the other side, the photonic transmitted power $P_{R_S-R_D}$ lowers with $n$ with a steeper rate at higher values of $T_S$ [$\mathcal{T}(\omega)\propto 1/n^2$]. The difference in the transmitted power between the clean and dirty limit predominantly arises from the difference in mobilities considered in the two cases ($\mu_{clean}=10^5~$cm$^2/$Vs and $\mu_{dirty}=10^4~$cm$^2/$Vs). At a fixed $n$, lower mobility implies larger values of impedance (resistance) of the reservoirs and, as a consequence, more effective power transfer between source and drain electronde [i.e. greater photonic transfer-function $\mathcal{T}(\omega)$].

\begin{figure}
\includegraphics {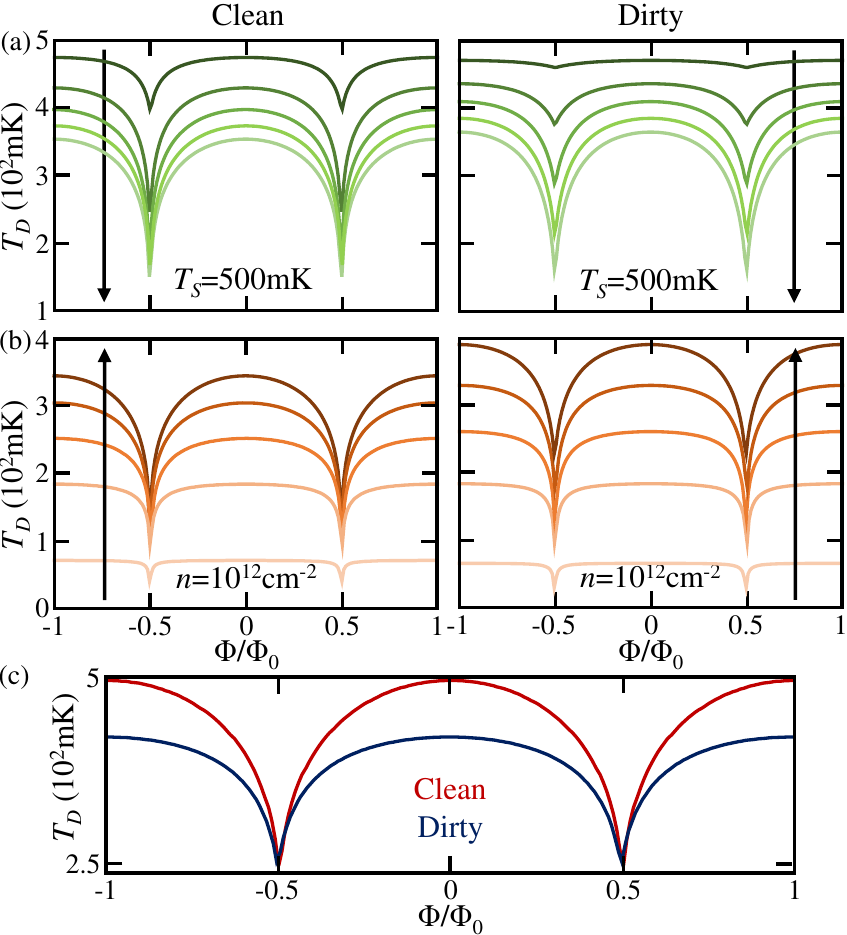}
\caption{\label{fig:Fig2mod2} (a) Drain temperature as a function of the magnetic flux $\Phi$ for $n=1,3,5,7,10\times 10^{11}$cm$^{-2}$ (rising in the arrow direction) at $T_S=500~$mK in the limits of clean (left) and dirty graphene (right). (b) Drain temperature as a function of $\Phi$ for source temperatures ranging from $100~$mK to $500~$mK with step of $100~$mK (as shown by the arrow) at $n=1\times 10^{12}$cm$^{-2}$ for clean (left) and dirty graphene (right). (c) Maximum achievable modulation of $T_D$ for clean (red) and dirty (blue) graphene.}
\end{figure}

Since $P_{R_S-R_D}(n)$ decreases for increasing $n$  (see Fig.~\ref{fig:Fig2mod1}), $T_{MAX}\doteq T_D(\Phi=0)$ decreases for increasing $n$ as shown in  Fig.~\ref{fig:Fig2mod2}-a in the case of $T_S=500~$mK. Since the heat current is not physically measurable, we evaluate the phase coherent modulation of thermal transport by monitoring $T_D$ as a function of the magnetic flux. The temperature modulation $\Delta T_D=T_D(\Phi=0)-T_D(\Phi=\Phi_0/2)$ increases with the charge carrier concentration of the electrodes for a fixed value of the source temperature both in the clean and the dirty limits. 

In Fig.~\ref{fig:Fig2mod2}-b we study $T_D$ as a function of $\Phi$ for different source temperatures at $n=1\times 10^{12}$cm$^{-2}$. The modulation is quite small for low temperatures (see lowest line in Figure~\ref{fig:Fig2mod2}-a representing $T_S=100~$mK), because the electron-phonon coupling is small and $P_{e-ph_{D}}\leq P_{R_S-R_D}$ for most values of $\Phi$. In the limit of clean graphene the maximum temperature modulation can be as large as $\Delta T_D\sim245~$mK at $T_S=950~$mK and $n=8\times 10^{11}$cm$^{-2}$, while in the dirty limit can be as large as $\Delta T_D\sim164~$mK at $T_S=550~$mK and $n=1\times 10^{12}cm^{-2}$ (see Fig.~\ref{fig:Fig2mod2}-c). The predicted temperature modulations are unprecedented even for galvanic heat transport in metallic systems \cite{Meschke2006, Giazotto2012, Fornieri2016}.

\begin{figure}
\includegraphics {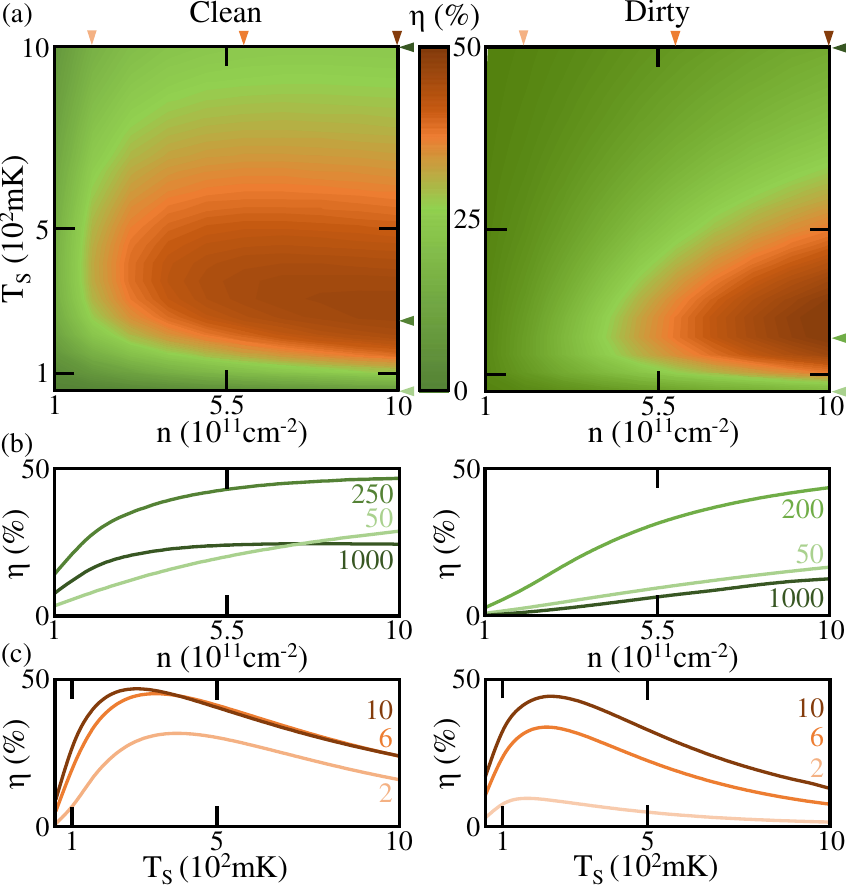}
\caption{\label{fig:Fig3} (a) Contour plot of $\eta$ as a function of $n$ and $T_S$. (b) $\eta$ vs $n$ for different values of $T_S$ (mK) indicated by the arrows in plot (a). (c) $\eta$ vs $T_S$ for different values of $n$ ($10^{12}$cm$^{-2}$) shown by the arrows in plot (a). Left (right) panel is for clean (dirty) graphene.}
\end{figure}

In order to envision the thermal efficiency of the system we introduce two parameters: the transmittance $\tau=T_D(\Phi=0)/T_S$ and the efficiency $\eta=(\Delta T_D/T_S)\times100$. $\tau$ decreases for increasing temperature, because the electron-phonon coupling becomes stronger [see Eq.~(\ref{eq:Pclean}) and Eq.~(\ref{eq:Pdirty})]. In particular, $\tau_{MAX}(T_S=50mK) \sim 0.99$ and  $\tau_{MAX}(T_S=1K) \sim 0.83$ in the clean limit, while $\tau_{MAX}(T_S=50mK) \sim 0.99$ and  $\tau_{MAX}(T_S=1K) \sim 0.92$ in the dirty limit. The efficiency parameter is shown in the contour plots of Fig.~\ref{fig:Fig3}-a as a function of $n$ and $T_S$. The behavior of $\eta$ is different in the case of clean (left panel of Fig.~\ref{fig:Fig3}-a) and dirty (right panel) graphene. In the first case, the efficiency of thermal conversion increases already for small $n$ and maintains high values for a wide range of temperatures with a maximum for $n=1\times~ 10^{12}cm^{-2}$ at $T_S=250~$mK. In the second case, $\eta$ increases gradually till large carrier concentrations with a reduced high conversion area characterized by a  maximum at  $n=1\times 10^{12}$cm$^{-2}$ and $T_S=200~$mK. Notably, in both cases the maximum value of $\eta$ is $\sim50\%$ and the system can work till high temperature (1K). This arises from the small values of the electron-phonon constant of graphene $\Sigma$ in all the parameters space. 

In order to discuss in detail the behavior of $\eta$ with $P_{R_S-R_D}$ and $P_{e-ph_{D}}$, we show some curves at fixed values of $T_S$ (Fig.~\ref{fig:Fig3}-b) and $n$ (Fig.~\ref{fig:Fig3}-c). At constant source temperature the efficiency increases with the carrier concentration, because the electron-photon coupling decreases more steeply near $\Phi=\Phi_0/2$ than near $\Phi=0$. In general, at high source temperature the device efficiency decreases, because the electron-phonon coupling is bigger. In the clean limit at high $T_S$, the difference between the maximum (at $\Phi=0$) and minimum (at $\Phi = \Phi_0/2$) value of $P_{R_S-R_D}$ starts to decrease and $P_{e-ph}$ rises. As a consequence, the efficiency decreases as shown by the curve at $T_S=1~$K of the left panel of Fig. \ref{fig:Fig3}-b. The increase of electron-phonon coupling with temperature [Eq.~(\ref{eq:Pclean}) and (\ref{eq:Pdirty})] is even more evident. In fact, all the curves in Fig.~\ref{fig:Fig3}-c show a non-monotonic behavior of the energy conversion efficiency. In the case of clean graphene all the maxima move towards low temperature by increasing carrier concentration, while for dirty graphene they shift in the opposite direction. This difference stems from the $T^4$ dependence of $P_{e-ph_{clean}}ì$ [see Eq.~(\ref{eq:Pclean})] and the $T^3$ of $P_{e-ph_{dirty}}$ [see Eq.~(\ref{eq:Pdirty})].

The system proposed in this work (see Fig.~\ref{fig:Fig1}-b) is the simplest arrangement able to coherently manipulate the thermal transport exclusively through tunable electron-photon and electron-phonon couplings. Increase the number of $SQUID$s in the coupling circuit would be the natural solution in order to improve the tunability of the electron-photon mediated thermal transport. On the other hand, it would complicate both the frequency-dependent transfer function and the experimental realization of the proposed device. Therefore, the choice of adding more SQUIDs needs to take in account the compromise between tunability and reliability. 

Finally, we want to point out that the system we propose can be also used to experimentally determine the dependence of the electron-phonon coupling in 2D electronic system with the charge carrier concentration, the defects concentration and the temperature \cite{Gasparinetti2011}. Within this approach the heat transferred to the sample under investigation can be controlled with unprecedented precision, thereby allowing to push the limit of the measurements to lower temperatures and higher accuracy.

\section{Conclusions}
In summary, we have studied the impact of electron-photon and electron-phonon coupling in the coherent heat transport between two bodies in non galvanic contact within a simple circuital approach. We have proposed a system consisting of a pair of gated 2DES capacitively-coupled through a dc-SQUID. This arrangement guarantees robustness on the device parameters of the performances and feasable nano-fabrication process. Furthermore, the architecture we propose ensures both high transmissivity and large modulation of thermal transport. In principle, the possibility to modulate the impedance of the source electrode could be used to control the temperature of another quantum device without intefeering with its behavior (by matching and un-matching the input and output impedances). In order to determine the efficiency of the system, we have presented quantitative results for source and drain electrodes made of graphene in the clean and dirty limits. In both cases, the proposed system shows unprecedented thermal modulations (up to $\Delta T_{D_{MAX}}\sim245~$mK), maximum transmittance $\tau_{MAX}\sim0.99$ and energy conversion efficiencies reaching $\eta_{MAX}\sim50\%$. The high modulation and control of thermal transport make this system the ideal platform for the investigation of electron-phonon coupling in 2D materials.

\begin{acknowledgments}
We  acknowledge  the  MIUR-FIRB2013-Project  Coca (Grant No. RBFR1379UX), the European Research Council under the European Unions Seventh Framework Programme (FP7/2007-2013)/ERC Grant No. 615187 - COMANCHE and the European Union (FP7/2007-2013)/REA Grant No. 630925 - COHEAT for partial financial support.
\end{acknowledgments}

\end{document}